\documentclass{PoS}

\title{Mixing angle of $K_1$ axial vector mesons}

\ShortTitle{Mixing angle of $K_1$ axial vector mesons}

\author{\speaker{Hai-Yang Cheng}\thanks{Work supported in part by the National Science Council of Taiwan, R.~O.~C. under Grant Nos.~NSC-100-2112-M-001-009-MY3. }\\
        Institute of Physics, Academia Sinica, Taipei, Taiwan 115, Republic of China\\
        E-mail: \email{phcheng@phys.sinica.edu.tw}}

\abstract{Analyses of various experimental measurements all indicate that the mixing angle $\theta_{K_1}$ of $K_1(1270)$ and $K_1(1400)$ is in the vicinity of $33^\circ$ or $57^\circ$. However, whether $\theta_{K_1}$ is greater  or less than $45^\circ$ is still quite controversial. For example, there were two very recent studies of  the strong decays of $K_1$ mesons. One group claimed that $\theta_{K_1}\approx 60^\circ$, while the other group obtained $\theta_{K_1}=(33.6\pm4.3)^\circ$. Since the determination of the mixing angles $\alpha_{^3\!P_1}$ and $\alpha_{^1\!P_1}$ with the former (latter) being the mixing angle of $f_1(1285)$ ($h_1(1170)$) and $f_1(1420)$ ($h_1(1380)$) in the flavor basis through mass relations depends on $\theta_{K_1}$, we show that $\theta_{K_1}\approx 57^\circ$ is ruled out as it leads to a too large deviation from ideal mixing in the $^1\!P_1$ sector, inconsistent with the lattice calculation of $\alpha_{^1\!P_1}$ and the observation of strong decays of $h_1(1170)$ and $h_1(1380)$.  We find that for $\theta_{K_1}\approx (28-30)^\circ$, the corresponding $\alpha_{^3\!P_1}$  and $\alpha_{^1\!P_1}$ agree well with all lattice and phenomenological analyses.  This again reinforces the statement that $\theta_{K_1}\sim 33^\circ$ is much more favored than $57^\circ$.

}

\FullConference{XV International Conference on Hadron Spectroscopy-Hadron 2013\\
		4-8 November 2013\\
		Nara, Japan }

\begin{document}

\def\be{\begin{eqnarray}}
\def\en{\end{eqnarray}}
\def\non{\nonumber}
\def\la{\langle}
\def\ra{\rangle}
\def\A{{\cal A}}
\def\B{{\cal B}}
\newcommand{\lsim}{\mathrel{\raisebox{-
.6ex}{$\stackrel{\textstyle<}{\sim}$}}}
\newcommand{\gsim}{\mathrel{\raisebox{-
.6ex}{$\stackrel{\textstyle>}{\sim}$}}}

\section{Introduction}
The mixing of the flavor-SU(3) singlet and octet states of vector and tensor
mesons to form mass eigenstates is of fundamental importance in hadronic
physics. According to the Appelquist-Carazzone
decoupling theorem, in a vectorial theory, as the mass of a particle gets large compared with a relevant scale, say, $\Lambda_{QCD} \simeq 300$ MeV, one can integrate this
particle out and define a low-energy effective field theory applicable below
this scale \cite{CS}.  Evidently, even though $m_s$ is not $\gg \Lambda_{QCD}$, there is
still a nearly complete decoupling for the case of vector mesons, namely, $\rho(770)$ and $\omega(892)$ states.  A similar situation of
near-ideal mixing occurs for the $J^{PC}=2^{++}$ tensor mesons
$f_2(1275)$, $f_2'(1525)$ and the $J^{PC}=3^{--}$  mesons
$\omega_3(1670)$, $\phi_3(1850)$ and this can also be understood in terms of approximate decoupling of the light $u \bar u + d \bar d$ state from the heavier $s \bar s$ state.

In the quark model, two nonets of $J^P=1^+$ axial-vector mesons are expected as the orbital excitation of the $q\bar q$ system.  In terms of the spectroscopic notation $^{2S+1}L_J$, there are two types of $P$-wave axial-vector mesons, namely, $^3\!P_1$ and $^1P_1$.  These two nonets have distinctive $C$ quantum numbers for the corresponding neutral mesons, $C=+$ and $C=-$, respectively.  Experimentally, the $J^{PC}=1^{++}$ nonet consists of $a_1(1260)$, $f_1(1285)$, $f_1(1420)$ and $K_{1A}$, while the $1^{+-}$ nonet contains $b_1(1235)$, $h_1(1170)$, $h_1(1380)$ and $K_{1B}$.  The non-strange axial vector mesons, for example, the
neutral $a_1(1260)$ and $b_1(1235)$ cannot have a mixing because of
the opposite $C$-parities. On the contrary, $K_{1A}$ and $K_{1B}$ are not the physical mass eigenstates $K_1(1270)$ and $K_1(1400)$ and they are mixed together due to the mass difference of strange and light quarks. Following the common convention we write
\be \label{K1mixing}
 \left( \begin{array}{c}
    |K_1(1270) \rangle \\
    |K_1(1400)   \rangle \end{array} \right ) =
\left( \begin{array}{cc}
     \sin\theta_{K_1} & \cos\theta_{K_1}   \\
   \cos\theta_{K_1}  & -\sin\theta_{K_1} \end{array} \right )
    \left( \begin{array}{c}
                 |K_{1A} \rangle \\
                 |K_{1B} \rangle \end{array} \right ) \ .
\en
Various phenomenological studies indicate that the $K_{1A}$-$K_{1B}$ mixing angle $\theta_{K_1}$ is around  either $33^\circ$ or $57^\circ$, \footnote{As discussed in \cite{Cheng:K1} and many early publications, the sign ambiguity of $\theta_{K_1}$ can be removed by fixing the relative sign of the decay constants of $K_{1A}$ and $K_{1B}$. We shall choose the convention of decay constants in such a way that $\theta_{K_1}$ is always positive.}
but there is no consensus as to whether this angle is greater or less than $45^\circ$.

We have shown in \cite{Cheng:K1} that the mixing angle $\theta_{K_1}$ can be pinned down based on the observation that when the $f_1(1285)$-$f_1(1420)$ mixing angle $\theta_{^3\!P_1}$ and the $h_1(1170)$-$h_1(1380)$ mixing angle $\theta_{^1\!P_1}$ are determined from the mass relations, they depend on the masses of $K_{1A}$ and $K_{1B}$, which in turn depend on $\theta_{K_1}$.
Since nearly ideal mixing occurs for vector, tensor and $3^{--}$ mesons except for pseudoscalar mesons where the axial anomaly plays a unique role, this feature is naively expected to hold  also for axial-vector mesons.
Lattice calculations of $\theta_{^1\!P_1}$ and the phenomenological analysis of the strong decays of $h_1(1170)$ and $h_1(1380)$ will enable us to discriminate the two different solutions for $\theta_{K_1}$. In this talk we will elaborate on this in more detail.

\section{Mixing of axial-vector mesons}

There exist several estimations on the mixing angle $\theta_{K_1}$ in the literature.  From the early experimental information on masses and the partial rates of $K_1(1270)$ and $K_1(1400)$, Suzuki found two possible solutions $\theta_{K_1}\approx 33^\circ$ and $57^\circ$ \cite{Suzuki}.  A similar constraint $35^\circ\lsim \theta_{K_1}\lsim 55^\circ$ was obtained in Ref.~\cite{Goldman} based solely on two parameters: the mass difference between the $a_1(1260)$ and $b_1(1235)$ mesons and the ratio of the constituent quark masses.  An analysis of $\tau\to K_1(1270)\nu_\tau$ and $K_1(1400)\nu_\tau$ decays also yielded the mixing angle to be $\approx 37^\circ$ or $58^\circ$ \cite{Cheng:DAP}.\footnote{Note that the mixing angle results in \cite{Cheng:DAP} based on CLEO \cite{CLEO} and OPEL \cite{OPEL} data differ from the the ones obtained in the CLEO paper \cite{CLEO}.}
Another determination of $\theta_{K_1}$ comes from the $f_1(1285)$-$f_1(1420)$ mixing angle $\theta_{^3\!P_1}$  to be introduced shortly below which can be reliably estimated from the analysis of the radiative decays $f_1(1285)\to \phi\gamma, \rho^0\gamma$ \cite{Close:1997nm}. A recent updated analysis yields $\theta_{^3\!P_1}=(19.4^{+4.5}_{-4.6})^\circ$ or $(51.1^{+4.5}_{-4.6})^\circ$ \cite{KCYang}.\footnote{From the same radiative decays, it was found $\theta_{^3\!P_1}=(56^{+4}_{-5})^\circ$ in \cite{Close:1997nm}. This has led some authors (e.g. \cite{DMLi}) to claim that  $\theta_{K_1}\sim 59^\circ$. However, another solution, namely, $\theta_{^3\!P_1}=(14.6^{+4}_{-5})^\circ$ corresponding to a smaller $\theta_{K_1}$, was missed in \cite{Close:1997nm}.
}
As we shall see below, the mixing angle $\theta_{^3\!P_1}$ is correlated to $\theta_{K_1}$. The corresponding $\theta_{K_1}$ is found to be $(31.7^{+2.8}_{-2.5})^\circ$ or $(56.3^{+3.9}_{-4.1})^\circ$. Therefore, all the analyses yield a mixing angle $\theta_{K_1}$ in the vicinity of either $33^\circ$ or $57^\circ$.

However, there is no consensus as to whether $\theta_{K_1}$ is greater or less than $45^\circ$.
It was found in the non-relativistic quark model that $m^2_{K_{1A}}<m^2_{K_{1B}}$ \cite{DMLi,Burakovsky,Chliapnikov} and hence $\theta_{K_1}$ is larger than $45^\circ$.
Interestingly, $\theta_{K_1}$ turned out to be of order $34^\circ$ in the relativized  quark model of \cite{Godfrey}.
Based on the covariant light-front model \cite{CCH}, the value of $51^\circ$ was found by the analysis of \cite{Cheng:2009ms}. From the study of $B\to K_1(1270)\gamma$ and $\tau\to K_1(1270)\nu_\tau$ within the framework of light-cone QCD sum rules, Hatanaka and Yang advocated that $\theta_{K_1}=(34\pm13)^\circ$ \cite{Hatanaka:2008xj}. There existed two recent studies of strong decays of $K_1(1270)$ and $K_1(1400)$ mesons with different approaches. One group obtained $\theta_{K_1}\approx 60^\circ$ based on the $^3P_0$ quark-pair-creation model for $K_1$ strong decays \cite{Kou}, while the other group found $\theta_{K_1}=(33.6\pm4.3)^\circ$ using a phenomenological flavor symmetric relativistic Lagrangian \cite{Giacosa}.
In short, there is a variety of different values of the mixing angle cited in the literature. It is the purpose of this work to pin down $\theta_{K_1}$.

We next consider the mixing of the isosinglet $1^3P_1$ states, $f_1(1285)$ and $f_1(1420)$, and the $1^1P_1$ states, $h_1(1170)$ and $h_1(1380)$ in the quark flavor and octet-singlet bases:
\begin{eqnarray}
 \left( \begin{array}{c}
    |f_1(1285) \rangle \\
    |f_1(1420)   \rangle \end{array} \right ) =
\left( \begin{array}{cc}
     \cos\theta_{^3\!P_1} & \sin\theta_{^3\!P_1}   \\
   -\sin\theta_{^3\!P_1}  & \cos\theta_{^3\!P_1} \end{array} \right )
    \left( \begin{array}{c}
                 |f_1 \rangle \\
                 |f_8 \rangle \end{array} \right )=
\left( \begin{array}{cc}
     \cos\alpha_{^3\!P_1} & \sin\alpha_{^3\!P_1}   \\
   -\sin\alpha_{^3\!P_1}  & \cos\alpha_{^3\!P_1} \end{array} \right )
    \left( \begin{array}{c}
                 |f_q \rangle \\
                 |f_s \rangle \end{array} \right )  \ ,
\end{eqnarray}
and
\begin{eqnarray}
 \left( \begin{array}{c}
    |h_1(1170) \rangle \\
    |h_1(1380)   \rangle \end{array} \right ) =
\left( \begin{array}{cc}
     \cos\theta_{^1\!P_1} & \sin\theta_{^1\!P_1}   \\
   -\sin\theta_{^1\!P_1}  & \cos\theta_{^1\!P_1} \end{array} \right )
    \left( \begin{array}{c}
                 |h_1 \rangle \\
                 |h_8 \rangle \end{array} \right ) =
\left( \begin{array}{cc}
     \cos\alpha_{^1\!P_1} & \sin\alpha_{^1\!P_1}   \\
   -\sin\alpha_{^1\!P_1}  & \cos\alpha_{^1\!P_1} \end{array} \right )
    \left( \begin{array}{c}
                 |h_q \rangle \\
                 |h_s \rangle \end{array} \right )       \ ,
\end{eqnarray}
where $f_1=(u\bar u+d\bar d+s\bar s)/\sqrt{3}$, $f_8=(u\bar u+d\bar d-2s\bar s)/\sqrt{6}$, $f_q=(u\bar u+d\bar d)/\sqrt{2}$, $f_s=s\bar s$ and likewise for $h_1$, $h_8$, $h_q$ and $h_s$. The mixing angle $\alpha$ in the flavor basis is related to the singlet-octet mixing angle $\theta$ by the relation $\alpha=35.3^\circ-\theta$. Therefore, $\alpha$ measures the deviation from ideal mixing.
Applying the Gell-Mann Okubo relations for the mass squared of the octet states
\be
m_8^2(^3\!P_1) &\equiv& m_{^3\!P_1}^2={1\over 3}(4m_{K_{1A}}^2-m_{a_1}^2), \qquad
m_8^2(^1\!P_1) \equiv m_{^1\!P_1}^2={1\over 3}(4m_{K_{1B}}^2-m_{b_1}^2),
\en
we obtain the following mass relations for the mixing angles $\theta_{^1\!P_1}$ and $\theta_{^3\!P_1}$ (for details, see \cite{Cheng:K1})
\begin{eqnarray} \label{tantheta:A}
\tan\theta_{^3\!P_1}&=& \frac{m_{^3\!P_1}^2-  m_{f'_1}^2} {\sqrt { m_{^3\!P_1}^2(m_{f_1}^2+m_{f'_1}^2- m_{^3\!P_1}^2)-m_{f_1}^2m^2_{f'_1}}}
                             \,, \nonumber\\
\tan\theta_{^1\!P_1}&=& \frac{ m_{^1\!P_1}^2-  m_{h'_1}^2} {\sqrt { m_{^1\!P_1}^2(m_{h_1}^2+m_{h'_1}^2-m_{^1\!P_1}^2)-m_{h_1}^2m^2_{h'_1}}}
                             \,,
\end{eqnarray}
where $f_1$ and $f'_1$ ($h_1$ and $h'_1$) are the short-handed notations for $f_1(1285)$ and $f_1(1420)$ ($h_1(1170)$ and $h_1(1380)$), respectively, and
\begin{eqnarray} \label{K1Amass}
 m_{K_{1A}}^2 &=& m_{K_1(1400)}^2 \cos^2\theta_{K_1} + m_{K_1(1270)}^2
 \sin^2\theta_{K_1} \,, \nonumber \\
  m_{K_{1B}}^2 &=&
 m_{K_1(1400)}^2 \sin^2\theta_{K_1} + m_{K_1(1270)}^2 \cos^2\theta_{K_1} \,.
\end{eqnarray}
It is clear that the mixing angles $\theta_{^3\!P_1}$ and $\theta_{^1\!P_1}$ depend on the masses of $K_{1A}$ and $K_{1B}$ states, which in turn depend on the $K_{1A}$-$K_{1B}$ mixing angle $\theta_{K_1}$. Table \ref{tab:axial} exhibits the values of $\alpha_{^3\!P_1}$, $\theta_{^3\!P_1}$ and $\alpha_{^1\!P_1}$, $\theta_{^1\!P_1}$ calculated using Eq. (\ref{tantheta:A}) for some representative values of $\theta_{K_1}$.

\begin{table}[t]
\caption{The values of the $f_1(1285)$-$f_1(1420)$ and $h_1(1170)$-$h_1(1380)$ mixing angles in the quark flavor (upper) and octet-singlet (lower) bases calculated using Eq. (2.4) for some representative $K_{1A}$-$K_{1B}$ mixing angle $\theta_{K_1}$.} \label{tab:axial}
\begin{center}
\begin{tabular}{|c| c c c c c c |}
\hline
~~$\theta_{K_1}$~~ & $57^\circ$ & $51^\circ$ & $45^\circ$ & $33^\circ$ & $30^\circ$ & $28^\circ$ \\
 \hline
 $\alpha_{^3\!P_1}$ & $16.5^\circ$ & $9.6^\circ$ & $2.4^\circ$ & $-13.7^\circ$ & $-18.9^\circ$ & $-23.5^\circ$ \\
 $\alpha_{^1\!P_1}$ & $-53.0^\circ$ & $-44.6^\circ$ & $-21.1^\circ$ & $-6.4^\circ$ & $-3.8^\circ$ & $-2.4^\circ$ \\
 \hline
 $\theta_{^3\!P_1}$ & $52^\circ$ & $45^\circ$ & $38^\circ$ & $22^\circ$ & $16^\circ$ & $12^\circ$ \\
 $\theta_{^1\!P_1}$ & $-18^\circ$ & $-9^\circ$ & $14^\circ$ & $29^\circ$ & $32^\circ$ & $33^\circ$ \\
 \hline
\end{tabular}
\end{center}
\end{table}

\section{Discussion}
We see from Table \ref{tab:axial} that the $K_{1A}$-$K_{1B}$ mixing angle
$\theta_{K_1}\approx 57^\circ$ corresponds to $\alpha_{^1P_1}=-53^\circ$ which is too far away from ideal mixing for the $^1P_1$ sector. Indeed, it is in violent disagreement with the lattice result $\alpha_{^1\!P_1}=\pm(3\pm1)^\circ$ obtained by the Hadron Spectrum Collaboration \cite{Dudek:2011tt}. Since only the modes $h_1(1170)\to\rho\pi$ and $h_1(1380)\to K\bar K^*,\bar KK^*$ have been seen so far, this implies that the quark content is primarily $s\bar s$ for $h_1(1380)$ and $q\bar q$ for $h_1(1170)$. Indeed, if $\theta_{K_1}=57^\circ$, we will have $h_1(1170)=0.60n\bar n-0.80s\bar s$ and $h_1(1380)=0.80n\bar n+0.60s\bar s$ with $n\bar n=(u\bar u+d\bar d)/\sqrt{2}$. It is obvious that the large $s\bar s$ content of $h_1(1170)$ and $n\bar n$ content of $h_1(1380)$ cannot explain why only the strong decay modes $h_1(1170)\to\rho\pi$ and $h_1(1380)\to K\bar K^*,\bar KK^*$ have been seen thus far. Therefore, it is evident that $\theta_{K_1}\approx 57^\circ$ is ruled out.

Can we conclude that $\theta_{K_1}$ is less than $45^\circ$ ? Let's examine the mixing angle $\alpha_{^3\!P_1}$. There are some information available. First, the radiative decay $f_1(1285)\to\phi\gamma$ and $\rho\gamma$ yields $\alpha_{^3\!P_1}=\pm(15.8^{+4.5}_{-4.6})^\circ$ \cite{KCYang}. An updated lattice calculation gives $\alpha_{^3\!P_1}=\pm(27\pm2)^\circ$ \cite{Dudek:2013yja}. A study of $B_{d,s}\to J/\psi f_1(1285)$ decays by LHCb leads to $\alpha_{^3\!P_1}=\pm(24.0^{+3.1+0.6}_{-2.6-0.8})^\circ$ \cite{LHCb}. Hence, $\alpha_{^3\!P_1}$ lies in the range $\pm (15\sim 27)^\circ$. Unlike the $^1P_1$ sector, the deviation of  $f_1(1285)$-$f_1(1420)$ mixing from the ideal one is sizable. Nevertheless,  the quark content is still primarily $s\bar
s$ for $f_1(1420)$ and $q\bar q$ for $f_1(1285)$. Indeed, $K^*\bar K$ and $K\bar K\pi$ are the dominant modes of
$f_1(1420)$ whereas $f_1(1285)$ decays mainly to the $\eta\pi\pi$ and $4\pi$
states.
It is clear from from Table \ref{tab:axial} that when $\theta_{K_1}\approx (28-30)^\circ$, the corresponding $\alpha_{^3\!P_1}$  and $\alpha_{^1\!P_1}$ agree well with all lattice and phenomenological analyses.  This in turn reinforces the statement that $\theta_{K_1}\sim 33^\circ$ is much more favored than $57^\circ$.

Two remarks are in order:  (i) The $K_1$ mixing angle $\theta_{K_1}\approx 57^\circ$ leads to acceptable $\alpha_{^3\!P_1}$ but too large $\alpha_{^1\!P_1}$. (ii) In the octet-singlet basis, the mixing angles are of order $\theta_{^3\!P_1}\sim 15^\circ$ and $\theta_{^1\!P_1}\sim 32^\circ$.

\section{Conclusions}
The $K_1$ mixing angle $\theta_{K_1}\approx 57^\circ$ is ruled out as it will lead to a too large deviation from ideal mixing in the $^1P_1$ sector, inconsistent with the observation of strong decays of $h_1(1170)$ and $h_1(1380)$ and a recent lattice calculation of $\theta_{^1\!P_1}$. We found when $\theta_{K_1}\approx (28-30)^\circ$, the corresponding $\alpha_{^3\!P_1}$  and $\alpha_{^1\!P_1}$ agree well with all lattice and phenomenological analyses.  This again implies that $\theta_{K_1}\sim 33^\circ$ is much more favored than $57^\circ$.


\begin{thebibliography}{99}

\bibitem{CS}  H.~Y.~Cheng and R.~Shrock,
  Phys.\ Rev.\ D {\bf 84}, 094008 (2011)
  [arXiv:1109.3877 [hep-ph]].


\bibitem{Cheng:K1}
  H.~Y.~Cheng,
  Phys.\ Lett.\ B {\bf 707}, 116 (2012)
  [arXiv:1110.2249 [hep-ph]].

\bibitem{Suzuki} M. Suzuki, Phys. Rev. D {\bf 47}, 1252 (1993).

\bibitem{Goldman} L. Burakovsky and T. Goldman, Phys. Rev. D {\bf 56}, 1368 (1997).

\bibitem{Cheng:DAP}
      H.~Y.~Cheng,
      Phys.\ Rev.\ D {\bf 67}, 094007 (2003).

\bibitem{Close:1997nm}
  F.~E.~Close and A.~Kirk,
  Z.\ Phys.\ C {\bf 76}, 469 (1997)
  [hep-ph/9706543].

\bibitem{KCYang}
  K.~C.~Yang,
  Phys.\ Rev.\  D {\bf 84}, 034035 (2011)
  [arXiv:1011.6113 [hep-ph]].

\bibitem{CLEO} D.M. Anser {\it et al.} [CLEO Collaboration], Phys. Rev. D {\bf 62}, 072006 (2000).

\bibitem{OPEL}
 G. Abbiendi {\it et al.} [OPAL Collaboration], Eur. J. Phys. C {\bf13}, 197 (2000).

\bibitem{DMLi}
  D.~M.~Li, B.~Ma and H.~Yu,
  Eur.\ Phys.\ J.\  A {\bf 26}, 141 (2005)
  [hep-ph/0509215];  D.~M.~Li and Z.~Li,
  Eur.\ Phys.\ J.\  A {\bf 28}, 369 (2006)
  [hep-ph/0606297].

\bibitem{Burakovsky}
  L.~Burakovsky and T.~Goldman,
  Phys.\ Rev.\  D {\bf 57}, 2879 (1998)
  [hep-ph/9703271].

\bibitem{Chliapnikov}
  P.V. Chliapnikov, Phys. Lett. B {\bf 496}, 129 (2000).

\bibitem{Godfrey}
  S.~Godfrey and N.~Isgur,
  Phys.\ Rev.\  D {\bf 32}, 189 (1985).

\bibitem{CCH}
    H.~Y.~Cheng, C.~K.~Chua and C.~W.~Hwang,
    Phys.\ Rev.\  D {\bf 69}, 074025 (2004)
    [arXiv:hep-ph/0310359].

\bibitem{Cheng:2009ms}
  H.~Y.~Cheng and C.~K.~Chua,
  Phys.\ Rev.\  D {\bf 81}, 114006 (2010)
  [arXiv:0909.4627 [hep-ph]].

\bibitem{Hatanaka:2008xj}
  H.~Hatanaka and  K.~C.~Yang,
  Phys.\ Rev.\  D {\bf 77}, 094023 (2008)
  [arXiv:0804.3198 [hep-ph]].

\bibitem{Kou}
  A.~Tayduganov, E.~Kou and A.~Le Yaouanc,
  Phys.\ Rev.\ D {\bf 85}, 074011 (2012)
  [arXiv:1111.6307 [hep-ph]].

\bibitem{Giacosa}
  F.~Divotgey, L.~Olbrich and F.~Giacosa,
  Eur.\ Phys.\ J.\ A {\bf 49}, 135 (2013)
  [arXiv:1306.1193 [hep-ph]].


\bibitem{Dudek:2011tt}
  J.~J.~Dudek, R.~G.~Edwards, B.~Joo, M.~J.~Peardon, D.~G.~Richards, and C.~E.~Thomas,
  Phys.\ Rev.\  D {\bf 83}, 111502 (2011)
  [arXiv:1102.4299 [hep-lat]].

\bibitem{Dudek:2013yja}
  J.~J.~Dudek, R.~G.~Edwards, P.~Guo and C.~E.~Thomas,
  arXiv:1309.2608 [hep-lat].

\bibitem{LHCb}
  R. Aaij {\it et al.}  [LHCb Collaboration],
  arXiv:1310.2145 [hep-ex].

\end{thebibliography}
\end{document}